\documentclass[prl,aps,amssymb,twocolumn,showpacs,superscriptaddress]{revtex4}

\usepackage{graphicx}

\begin{document}

\title{Noise Thermal Impedance of a Diffusive Wire}

\author{B.~Reulet}
\affiliation{Departments of Applied Physics and Physics, Yale University, New Haven CT 06520-8284, USA}
\affiliation{Laboratoire de Physique des Solides, UMR8502, b\^at 510, Universit\'e Paris-Sud 91405 Orsay, France}
\author{D.E.~Prober}
\affiliation{Departments of Applied Physics and Physics, Yale University, New Haven CT 06520-8284, USA}

\date{\today}

\begin{abstract}

The current noise density $S_2$ of a conductor in equilibrium, the Johnson noise, is determined by its temperature $T$: $S_2=4k_BTG$ with $G$ the conductance. The sample's noise temperature $T_N=S_2/(4k_BG)$ generalizes $T$ for a system out of equilibrium. We introduce the "noise thermal impedance" of a sample as the amplitude of the oscillation of $T_N$ when heated by an oscillating power. For a macroscopic sample, it is the usual thermal impedance. We show for a diffusive wire how this (complex) frequency-dependent quantity gives access to the electron-phonon interaction time in a long wire and to the diffusion time in a shorter one, and how its real part may also give access to the electron-electron inelastic time. These times are not simply accessible from the frequency dependence of $S_2$ itself.
\end{abstract}
\pacs{72.70.+m, 05.40.-a, 73.23.-b, 07.57.Kp}
\maketitle

For decades, measurements of the electrical response of condensed matter systems have provided powerful insights into the physics at the macro- and micro-scale.  Measurements of the conductance $G=dI/dV$ (with $I$ the current when biased by a voltage $V$), and of the noise (the variance of the current fluctuations) are examples of the many successful approaches\cite{BuBlan}.  Yet, for good conductors, the frequency dependence of the conductance and the noise is determined only by charge screening by the electron fluid, due to the long range interaction of the electrons.  For normal metals (non-superconductors), the physics associated with inelastic processes, energy exchange, dephasing, or diffusion times, is usually accessible only in the quantum corrections\cite{qcorrections} or from tunneling measurements on specific materials\cite{saclay}. Superconductors are an exception since the gap, and near $T_C$ the resistance, is sensitive to the distribution of excitations. Thus, time- or frequency-domain transport measurements provide \textit{direct} access to the time scales of microscopic processes, such as electron-phonon inelastic relaxation\cite{gersh} , diffusion removal of energy \cite{HEBs}, or quasiparticle recombination\cite{recomb}.

        In this letter we develop the theory for a novel approach to directly measure dynamic processes of electrons in a normal metal.  For an electron system in equilibrium ($V=0$), the temperature is reflected in the Fermi-Dirac distribution of state occupancy and can be determined from the Johnson-Nyquist noise.  If the occupancy is perturbed in a charge-neutral fashion, its relaxation is governed by the microscopic processes that we wish to access.  One can determine the relaxation of the electron temperature (and more generally, excitations that are charge-neutral) from the time-dependence of the magnitude of the noise (measured at frequencies much higher than the inverse relaxation time) when the system is driven by an ac voltage. Our idea is, in essence, to use the driven noise to determine the dynamics. This is closely related to the recently proposed and explored third moment of the current noise\cite{Pilgram,S3nous}.

We consider a conductor biased by a time-dependent voltage  $V(t)=V_{dc}+\delta V\cos \omega t$. For simplicity we treat $e\delta V\ll eV_{dc},k_BT$. The mean square current fluctuations are measured through the spectral density of the current noise, $S_2$, integrated over a frequency band $\Delta\Omega$ around the frequency $\Omega$. In equilibrium ($V(t)=0$), $S_2=4k_BTG$ with $G$ the electrical conductance, taken to be independent of $T$ and $V$\cite{footG}. $S_2$ is averaged over a time $\tau_m$ such that $\omega\ll\tau_m^{-1}\ll\Delta\Omega,\Omega$, to give $S_2(t)$. Experimentally, this could be implemented by coupling the sample noise through a bandpass filter centered at frequency $\Omega$ to a bolometer with a response time $\tau_m$ \cite{bolo}. We treat low-frequency  noise, $\hbar\Omega\ll eV_{dc},k_BT$, so our conclusions do not depend on $\Omega$, $\Delta\Omega$ or $\tau_m$.  Under the time-dependent bias $V(t)$, $S_2(t)$ is amplitude modulated at frequency $\omega$ (see inset, Fig. 2). We define the sample's noise temperature $T_N(t)=S_2(t)/(4k_BG)$, and the instantaneous Joule power $P_J=I(t)V(t)=GV^2(t)$ dissipated in the sample. From their (complex) components $\delta T_N^\omega$ and $\delta P_J^\omega$ at frequency $\omega$, one defines the (complex) response function $\mathcal{R}(\omega)=\delta T_N^\omega/\delta P_J^\omega$. $\mathcal{R}$ measures how much the noise temperature oscillates when the system is heated by an oscillating power. $\mathcal{R}$ has units of a thermal resistance, K/W. For a macroscopic sample, $T_N$ is the sample temperature, and $\mathcal{R}$ is simply the thermal impedance between the sample and its environment. Thus, we will call $\mathcal{R}(\omega)$ the "noise thermal impedance" (NTI) of the sample at frequency $\omega$. It exhibits a frequency dependence on the scale of the inverse thermal relaxation time. For a thin film or wire at low temperature, as we consider later, this thermalization time is determined by energy removal  processes experienced by the electrons (electron-hole relaxation)\cite{foot1}.

In the following we calculate $\mathcal{R}(\omega)$ for a diffusive wire of length $L$ between two normal metal reservoirs (see inset of Fig. 1), in several limiting cases: i) long wires $L_{e-e}\ll L_{e-ph}\ll L$; ii) wires of intermediate length $L_{e-e}\ll L\ll L_{e-ph}$; and iii) short wires $L\ll L_{e-e}$. Here $L_{e-ph}$ stands for the electron-phonon interaction length and $L_{e-e}$ for the energy relaxation length due to electron-electron (e-e) interaction. These lengths, much longer than the mean free path, are related to the corresponding times by, e.g., $L_{e-e}^2=D\tau_{e-e}$ with $D$ the diffusion coefficient. We show that $\mathcal{R}(\omega)$ gives access to the electron-phonon relaxation time in i), and that it gives the diffusion time $\tau_D=L^2/D$ for ii) and iii). The two latter cases differ significantly if one measures the real and imaginary parts of $\mathcal{R}(\omega)$. Using this difference one can probe the electron-electron interaction time. Finally we discuss how $\mathcal{R}(\omega)$ is related to the third cumulant of the noise and its environmental corrections.

A general properties of $\mathcal{R}(\omega)$ is that, at low frequency, the magnitude of the noise follows adiabatically the voltage variations, such that: $\mathcal{R}(\omega=0)=dT_N/dP_J$. The Joule power has a component at frequency $\omega$, $\delta P_J^\omega=2GV_{dc}\delta V$, so one has $\mathcal{R}(0)=(dS_2/dV)/(8k_BG^2V_{dc})$ \cite{foot2}.

i) We first consider a long wire, $L\gg L_{e-ph}$. The electrons give the energy they acquire from the electric field to the phonons. We refer to this regime as phonon-cooled. For a wire made of a thin film, the phonons of the film and substrate are well coupled and represent the thermal bath\cite{gersh}. Phonon emission occurs uniformly in the wire, except near the ends, on a length $L_{e-ph}$ where the hot electrons can leave the sample without emitting a phonon. Such finite length effects  are negligible for $L\gg L_{e-ph}$, so we consider the electron temperature $T_e(t)=T_N(t)$ to be position independent. In the absence of ac excitation, $T_e(T,V_{dc})$ is such that the phonon cooling power $P_{e-ph}(T_e,T)$ equals the Joule power $GV_{dc}^2$. The electron-phonon thermal conductance $G_{e-ph}=dP_{e-ph}/dT_e$ has been studied with dc heating\cite{Roukes85}. For  ac excitation we have:
\begin{equation}
C_e(t)\frac{dT_e}{dt}=P_J(t)-P_{e-ph}(T_e,T)
\end{equation}
where the phonons remain at temperature $T$ \cite{foot1}; $C_e=\gamma T_e$ is the electron heat capacity. The electron temperature oscillates: $T_e(t)=T_e(V_{dc})+\mathrm{Re}[\delta T_e^\omega\exp(i\omega t)]$, and:
\begin{equation}
\mathcal{R}(\omega)=\frac{\delta T_e^\omega}{\delta P_J^\omega}=
\frac{G_{e-ph}^{-1}}{(1+i\omega\tau_{e-ph})}
\label{eqLor}
\end{equation}
with $\tau_{e-ph}=C_e/G_{e-ph}$ the electron-phonon time at $T_e$. $\mathcal{R}(\omega)$ is the electron-phonon thermal impedance at temperature $T_e(T,V_{dc})$. Measurements of $T_N(t)$ for a voltage step have recently been undertaken \cite{ming}.

ii) We now turn to the case of intermediate length, $L_{e-e}\ll L\ll L_{e-ph}$. This is the hot electron, diffusion-cooled regime. The energy stored in the sample relaxes because energetic electrons leave the sample and are replaced by thermalized ones coming from the reservoirs. This occurs on a time scale set by the diffusion time $\tau_D$. One can still define a local temperature $T_e(x,t)$ since the electrons equilibrate with each other locally. $T_e$ is peaked along the wire, given by:
\begin{equation}
C_e(x,t)\frac{\partial T_e}{\partial t}=P_J(t)
+\frac{\partial}{\partial x}
\left( G_{WF}(x,t)\frac{\partial T_e}{\partial x}\right)
\label{eqhe}
\end{equation}
with the boundary conditions: $T_e(0,t)=T_e(1,t)=T$, with $T$ the temperature of the contacts. $x$ denotes the position along the wire in units of $L$: $0\leq x\leq1$.  $G_{WF}$ is the electron thermal conductance, related to the electrical conductance $G$ through the Wiedemann-Franz law: $G_{WF}=\mathcal{L}GT_e$ with $\mathcal{L}=(\pi^2/3)(k_B/e)^2$ the Lorentz number. Eq. (\ref{eqhe}) is linear in $T_e^2$ and we compute:
\begin{equation}
T_e^2(x,t)=T_0^2(x)+2\mathrm{Re}[A(x,\omega)\exp(i\omega t)]
\end{equation}
where $T_0$ is solution of the dc case, $T_0^2(x)=T^2(1+\alpha x(1-x))$ with $\alpha=(3/\pi^2)(eV_{dc})^2/(k_BT)^2$, and $A$ the ac solution of the $V_{dc}=0$ case (usually called the "weak heating" limit), for which $T_0=T$. We find:
\begin{equation}
A(x,\omega)=\frac{\delta P_J^\omega T}{G_{WF}q^2}
\left(1-\frac{\cosh q(x-1/2)}{\cosh q/2}\right)
\end{equation}
with $q=\sqrt{i\omega\tau_D}$. For a small ac excitation and $V_{dc}$ finite, the ac response of the electron temperature is given by: $\delta T_e^\omega(x)=A(x,\omega)/T_0(x)$. $T_N$ is the average of $T_e$ along the wire. For $eV_{dc}\ll k_BT$, $T_0(x)\simeq T$ and we obtain:
\begin{equation}
\frac{\mathcal{R}(\omega)}{\mathcal{R}(0)}=12\frac{q-2\tanh(q/2)}{q^3}
\end{equation}
and $\mathcal{R}(0)=G_{WF}^{-1}/12$. We do not have an analytical expression for $\mathcal{R}(\omega)$ for all $V_{dc}$, but numerical calculations show that the dependence of $\mathcal{R}(\omega)/\mathcal{R}(0)$ on $V_{dc}$  is extremely weak. Curves for different values of $V_{dc}$ are indistinguishable on a linear plot. Real and imaginary parts of $\mathcal{R}(\omega)$ as a function of $\omega\tau_D$ are plotted in Fig. 1. At high frequency, $\omega\tau_D\gg1$, $\mathrm{Re}[\mathcal{R}(\omega)]$ decays like $\omega^{-3/2}$ whereas $\mathrm{Im}[\mathcal{R}(\omega)]$ decays like $\omega^{-1}$. The magnitude $|\mathcal{R}(\omega)|$ is plotted on Fig. 2. The frequency for which $|\mathcal{R}(\omega)|^2=1/2$, i.e., the bandwidth of this "thermal" response, is $\sim10\tau_D^{-1}$.

\begin{figure}
\includegraphics[width= 0.9\columnwidth]{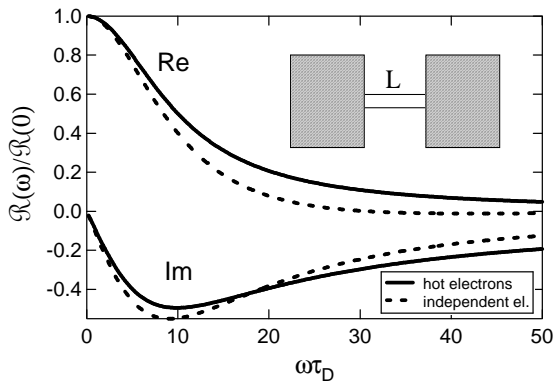}
\caption{Real and imaginary parts of $\mathcal{R}(\omega)$ as a function of $\omega\tau_D$, in the hot electron, diffusion cooled regime ii) (solid lines) and independent electron regime iii) (dashed lines). The values of $\mathcal{R}(0)$ differ by $\sim10$\% for $eV_{dc}\ll k_BT$. Inset: the geometry considered: a wire of length $L$ between two thick normal metal contacts.
}\end{figure}

iii) We now consider the case of elastic transport, $L\ll L_{e-e}$. This is the  independent-electron regime, since the electrons travel along the wire without experiencing  inelastic collisions. There is no local temperature, but one can define a local noise temperature:
\begin{equation}
T_N(x,t)=\int_{-\infty}^{+\infty} f(x,E,t)(1-f(x,E,t))dE/k_B
\label{eqTN}
\end{equation}
where $f(x,E,t)$ stands for the local energy distribution function in the wire. If $f$ is a Fermi function at temperature $T$, Eq. (\ref{eqTN}) gives $T_N=T$. The wire's noise temperature $T_N(t)$ is the average of $T_N(x,t)$ along the wire.
The distribution function $f(x,E,t)$ obeys the 1D diffusion equation\cite{footac}:
\begin{equation}
\frac{\partial f(x,E,t)}{\partial t}=
\frac{D}{L^2}\frac{\partial^2f(x,E,t)}{\partial x^2}
\label{eqgen}
\end{equation}
The effect of the external voltage appears only in the boundary conditions: $f(0,E,t)=f_F(E)$ and $f(1,E,t)=f_F(E+eV(t))$ with $f_F(E)$ the Fermi distribution function at temperature $T$. Solving Eq.(\ref{eqgen}) for $f$ to  first order in $\delta V$ we obtain the $\omega$ component of the time-dependent noise temperature profile: $T_N^\omega(x)\propto(1-x)\sinh(qx)/\sinh q$ with $q=\sqrt{i\omega\tau_D}$. We deduce:
\begin{equation}
\frac{\mathcal{R}(\omega)}{\mathcal{R}(0)}=6\frac{\sinh q-q}{q^2\sinh q}
\end{equation}
For $eV_{dc}\ll k_BT$ one has $\mathcal{R}(0)=(\pi^2/108)G_{WF}^{-1}\sim G_{WF}^{-1}/10.9$. This differs from case ii) by only $\sim10$\%. Real and imaginary parts of $\mathcal{R}(\omega)$ as a function of $\omega\tau_D$ are plotted on Fig. 1. For $\omega\tau_D\gg1$, $\mathrm{Re}[\mathcal{R}(\omega)]$ decays exponentially whereas $\mathrm{Im}[\mathcal{R}(\omega)]$ decays like $\omega^{-1}$. The magnitude $|\mathcal{R}(\omega)|$ is plotted on Fig. 2. The frequency for which $|\mathcal{R}(\omega)|^2=1/2$ is $\sim9\tau_D^{-1}$.

\begin{figure}
\includegraphics[width= 0.9\columnwidth]{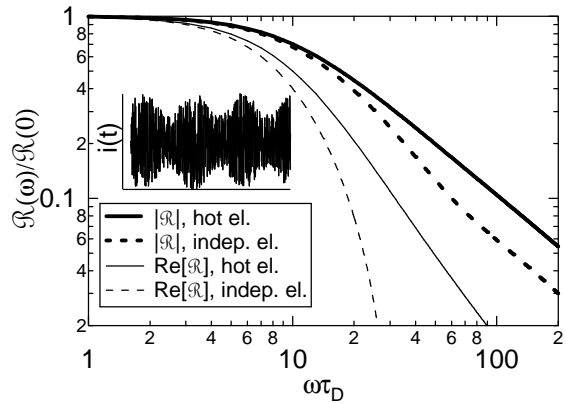}
\caption{Magnitude (thick lines) and real parts (thin lines) of $\mathcal{R}(\omega)$ as a function of $\omega\tau_D$, in the hot electrons, diffusion-cooled regime ii) (solid lines) and independent electrons regime iii) (dashed lines). Inset: Current noise amplitude-modulated by the time-dependent bias $V(t)$.
}\end{figure}

We now compare the three cases. $\mathcal{R}(\omega)$ has a Lorentzian shape, Eq. (\ref{eqLor}), for case i), and in ii) has a frequency dependence that is very similar.  The roll-off frequency, $\tau_{e-ph}^{-1}$, of $\mathcal{R}(\omega)$ for phonon cooling i) is temperature-dependent, since $\tau_{e-ph}\propto T^{-p}$, whereas the roll-off frequency for diffusion cooling ii) is related only to the diffusion time and is thus temperature-independent. A measurement of the roll-off frequency of $\mathcal{R}(\omega)$ vs. temperature (or dc voltage) in case i) gives a direct measurement of $\tau_{e-ph}(T_e)$. The shape of the magnitude $|\mathcal{R}(\omega)|$ in cases ii) and iii) is similar although case iii) exhibits a kink (see Fig. 2). However the real part of $\mathcal{R}(\omega)$ is quite different. For the independent electron regime iii), $\mathrm{Re}[\mathcal{R}(\omega)]$ crosses zero and is negative above  $\omega\tau_D\sim31$. At this frequency, in the hot electron, diffusion-cooled regime ii)  $\mathrm{Re}[\mathcal{R}(\omega)]/\mathcal{R}(0)$ has only dropped to $\sim0.1$. This remarkable difference is definitely measurable. Note that there is no principle preventing $\mathrm{Re}[\mathcal{R}]$ from being negative. This occurs when the average distribution function in the center of the wire oscillates out of phase with the excitation voltage.

The case of intermediate electron-electron time, $L\sim L_{e-e}$, is beyond the scope of this article. But this could be considered by adding e-e relaxation to the right hand side of Eq. (\ref{eqgen}). This eq., in the limit $\tau_{e-e}\rightarrow0$ leads to the heat diffusion equation (\ref{eqhe}) \cite{nagaev95}. It would be of interest to calculate how the existence of the zero of $\mathrm{Re}[\mathcal{R}(\omega)]$ in case iii) changes with finite e-e strength. Since e-e relaxation is energy-dependent, the position of the zero should be voltage- and temperature-dependent.

In this last section, we consider the relation of our results to other kinds of noise measurements, and other possible applications. As a first example of its relation, we can contrast the NTI we have calculated to the noise under ac voltage excitation considered previously, the so-called "photon assisted noise" \cite{thPAT,robPAT,Shytov,Pedersen}. The latter refers to the effect of an ac voltage on the \textit{time-averaged} noise. It has features at $eV_{dc}=n\hbar\omega$ (with $n$ integer). This differs significantly from the NTI, which measures the time dependence of the noise averaged on a time scale $\tau_m$, revealing the dynamics of the energy exchanges.

Our result also elucidates the importance of correlations in the scattering matrix formalism, which has been very powerful in treating noise properties of coherent systems\cite{BuBlan}. $\mathcal{R}(\omega)$ could be treated within this formalism, starting from its definition in terms of the classical fluctuating current at frequency $\Omega$, $i(\Omega)$ \cite{footorder}:
\begin{equation}
\mathcal{R}(\omega)=\frac{\langle i(\Omega)i(\omega-\Omega)\delta V_{-\omega}\rangle}
{8k_BG^2V_{dc}\langle|\delta V_\omega|^2\rangle}
\label{defr}
\end{equation}
The frequency dependence of $\mathcal{R}(\omega)$ on the scale of $\tau_D^{-1}$ for cases ii) and iii) comes only from correlations within the scattering matrix at different energies, on the scale of the Thouless energy $\hbar/\tau_D$. As a consequence, $\mathcal{R}$ provides a direct probe of the correlations, which are not considered in usual calculations of $S_2$.

Our calculation sheds new light on the environmental effects on the third cumulant of noise. These have been considered recently\cite{Kindermann,S3nous}, specifically for tunnel junctions. The expression of $\mathcal{R}(\omega)$ in Eq.(\ref{defr}) makes it appear as a third order correlation, like the third cumulant of the noise $S_3(\omega_1,\omega_2)=\langle i(\omega_1)i(\omega_2-\omega_1)i(-\omega_2)\rangle$, except the sample-generated current fluctuation at frequency $-\omega_2$ has been replaced by the external applied current (or voltage). As a matter of fact, it has been calculated that the external current noise, by modulating the noise emitted by the sample, does contribute to $S_3$ \cite{Kindermann}. This mechanism has been explicitly demonstrated in experiment by applying an ac voltage to a tunnel junction and detecting $S_3$ \cite{S3nous}. This extrinsic contribution to $S_3$ is $\propto\int d\omega S_2^{env}(\omega)\mathcal{R}(-\omega)$ with $S_2^{env}(\omega)$ the noise emitted by the environment; we suppose here $\hbar\omega\ll eV,k_BT$.

Our approach for the NTI can also be used for the calculation of the environmental effects on $S_3$ for a diffusive wire. In particular, we predict that the contribution of the environmental noise to $S_3$ vanishes at frequencies much larger than $\tau_D^{-1}$.
The intrinsic contributions to $S_3$ of a diffusive wire also decays for frequencies $>\tau_D^{-1}$, even for voltage bias\cite{Pilgram}. We believe this also may be understood from the behavior of $\mathcal{R}(\omega)$. Certainly the measurement of  $\mathcal{R}(\omega)$ is simpler than that of $S_3$.

The frequency scale of $\mathcal{R}(\omega)$ in cases ii) and iii) is set by the escape time of the electron-hole excitations  from the wire. We believe this statement applies qualitatively to other systems. Indeed, in  chaotic cavities, $\mathcal{R}(\omega)$ should also decay on the scale of the inverse dwell time (as $S_3$ does \cite{S3chaos}); in a quasi-ballistic wire, it likely decays on the scale of the inverse transit time $L/v_F$ with $v_F$ the Fermi velocity.
The use of our method for a carbon nanotube may provide an example of its applicability.  For most single wall nanotubes, it is not known if the conductance results from scattering that is equal for all four quantum channels, or from some of these channels being blocked, and the others open.  The time scale determined from $\mathcal{R}(\omega)$, might distinguish these two cases. The noise relaxation time scale of $\mathcal{R}(\omega)$ for a normal metal wire between superconducting reservoirs should also be studied. We suspect the relaxation time for this case is much longer than $\tau_D$, because electron-hole excitations of energy smaller than the superconducting gap of the reservoirs cannot escape the wire.

Last, we consider a generalization of our result. $\mathcal{R}(\omega)$ is a response to an external excitation. At equilibrium (for $V_{dc}=0$), it must be related to some correlation function through the fluctuation-dissipation theorem. The latter gives for a macroscopic system \cite{bolo} : $\langle\delta T^2(\omega)\rangle=4k_B\mathrm{Re}[G_{th}^{-1}(\omega)]T^2$. A similar expression should relate $\langle\delta T_N^2(\omega)\rangle$ to $\mathcal{R}(\omega)$. Since $T_N\propto S_2$, $\langle\delta T_N^2(\omega)\rangle$ is related to  $S_4$, the fourth cumulant of noise \cite{KimPil}. Thus, we conjecture that $\mathcal{R}$ is related to $S_4$.

We thank W. Belzig, M. B\"uttiker, M. Devoret, L. Levitov, S. Pilgram, M. Reese, P. Samuelsson, D. Santavicca, R. Schoelkopf, M. Shen and A. Shytov for fruitful discussions. This work was supported by NSF DMR-0407082.


\begin{thebibliography}{99}
\bibitem{footG} For $\omega\tau_{RC}\ll1$ (with $\tau_{RC}$ the $RC$ time), the admittance is the dc conductance $G$.
\bibitem{foot1} We will consider thin enough films at low enough temperature so we can neglect the Kapitza resistance between the phonons of the wire and the phonons of the substrate.
\bibitem{foot2} At zero voltage $V_{dc}=0$, the Joule power oscillates at $2\omega$. Similarly the noise, which is an even function of voltage, is proportional to $\delta V^2$ and oscillates at $2\omega$. Thus one simply has to replace $\omega$ by $2\omega$ in our definition to extend it to the case of zero dc bias.
\bibitem{footorder} In Eq. (\ref{defr}), the issue of ordering the current operators has been disregarded, for the sake of simplicity.
\end{thebibliography}
\end{document}